\documentclass[prb,preprint, superscriptaddress, nolongbibliography]{revtex4-2}

\usepackage[dvipdfmx]{graphicx}
\usepackage{dcolumn}
\usepackage{bm}

\usepackage{color}

\begin{document}


\title{Growth of 2D topological material Bi on InSb(111)B with fractal surface structures}

\author{Yoshiyuki Ohtsubo}
\email{y\_oh@qst.go.jp}
\affiliation{National Institutes for Quantum Science and Technology, Sendai 980-8579, Japan}
\affiliation{Graduate School of Frontier Biosciences, Osaka University, Suita 565-0871, Japan}
\affiliation{Department of Physics, Graduate School of Science, Osaka University, Toyonaka 560-0043, Japan}
\author{Takuto Nakamura}
\affiliation{Graduate School of Frontier Biosciences, Osaka University, Suita 565-0871, Japan}
\affiliation{Department of Physics, Graduate School of Science, Osaka University, Toyonaka 560-0043, Japan}
\author{Shin-ichi Kimura}
\email{kimura.shin-ichi.fbs@osaka-u.ac.jp}
\affiliation{Graduate School of Frontier Biosciences, Osaka University, Suita 565-0871, Japan}
\affiliation{Department of Physics, Graduate School of Science, Osaka University, Toyonaka 560-0043, Japan}
\affiliation{Institute for Molecular Science, Okazaki 444-8585, Japan}


\date{\today}

\begin{abstract}
Bismuth (Bi) atomic layers are known as 2D topological materials with variety of the electronic structures and topological orders depending on the number of stacking layers.
Recently, it is reported that few layers of Bi grown on semiconductor substrate InSb(111)B exhibit the Sierpi\'{n}ski-triangle (ST) fractal patterns on the surface.
In this work, we have grown Bi layers on InSb(111)B and traced the evolution of the atomic and electronic structures of Bi.
The surface atomic structures and growth modes were monitored by using reflective high-energy electron diffraction and core-level photoelectron spectroscopy.
It is suggested that the single layer of the ST-phase Bi grows on InSb(111)B and the following Bi deposition causes layer-by-layer growth up to nominally 4 atomic layers.
Diffuse band dispersion and quantum well states observed by angle-resolved photoelectron spectroscopy are consistent with the small surface domains and variation of the thickness even during the layer-by-layer growth region.
The further Bi evaporation changes the growth mode to the 3D island formation.
The unveiled growth behavior of Bi on InSb(111)B would provide a new interesting playground to study 2D topological electronic structure of quasi-periodic 2D atomic layers.
\end{abstract}

\maketitle

\section{Introduction}
Bismuth (Bi) is one of the most widely used elements as a component of topological materials \cite{Hasan10} such as the 3D topological insulators (TIs) Bi$_2$Se$_3$ \cite{Zhang09, Xia09} and Bi$_{1-x}$Sb$_x$ \cite{Hsieh08}, because it is the heaviest non-radioactive element with a strong spin-orbit interaction.
Not only the 3D crystals, 2D atomic layers of Bi(111) with buckled honeycomb lattice is also expected as a 2D TI with fertile topological phases depending on its thickness \cite{Murakami06, Liu11}.
Moreover, it is also expected that some external perturbations, such as lattice distortion, external fields, or point defects, could modify and control the topological phase and electronic structure of the 2D Bi layers \cite{Huang13, Kadioglu17, Liu17}.

From experimental aspect, the atomic and electronic structure and its topology of the 2D Bi(111) layers have been studied on various substrates.
The Bi atomic layers down to 6 atomic layers are known to grow on semiconductor Si(111) and Ge(111), while the thinner Bi(111) layers are not obtained because of the structural phase transition to phosphorene-like Bi(110) layers \cite{Nagao04, Hirahara06, Hatta09, Ito16}.
Note that another unit, bi-atomic layer (BL), is used in some references.
On a TI with close in-plane lattice constant, Bi$_2$Te$_3$, Bi(111) layers grow from 1 BL \cite{Hirahara11, Hirahara12, Chen12, Yao16}.
However, the 2D electronic structure of Bi(111) coexists with topological surface states of the substrate, Bi$_2$Te$_3$ \cite{Hirahara11}.
Although the single-layer Bi (Bismuthene) on SiC is free from substrate electronic structure around the Fermi level ($E_{\rm F}$), its atomic structure is strongly influenced by the substrate lattice, resulting in the flat Bi layer in contrast to the buckled ones in the free-standing case \cite{Reis17}.
Therefore, to fabricate the few atomic layers (typically 1-4) of free-standing 2D Bi(111) was an unresolved problem of 2D material growth, in order to explore its topological electronic structures.

Recently, epitaxial growth from 4 layers of Bi(111) was reported on InAs(111)A; the (111) surfaces of zinc-blende structures are often called as A (B) face with group-III (V) element termination \cite{Nicolai19}.
Although the detailed evolution of Bi 2D electronic structures are not reported yet, the zinc-blende (111) surfaces are spotlighted as promising substrate for growth of few layers of 2D Bi(111) thanks to their semiconducting electronic structures and triangular lattice with close values of the in-plane lattice constants to Bi(111); 4.53 \AA, 4.28 \AA\ and 4.58 \AA\ for Bi(111), InAs(111) and InSb(111), respectively.
Moreover, the other attractive result is reported in this year that Bi on InSb(111)B grows with Sierpi\'{n}ski-triangle (ST)-type fractal surface patterns by Scanning Tunneling Microscopy (STM) and theoretical model calculations \cite{Liu21}.
ST is a self-similar fractal with a fractional Hausdorff dimension \cite{Sierpinski15}.
This exotic surface structure provides an attractive playground to study the topological electronic structures in 2D quasicrystals.
For further studies of this exotic surface structure, experimental exploration about the growth modes and surface electronic structures of Bi layers on InSb(111)B is awaited.

In this work, we have traced the evolution of the atomic and electronic structures of the Bi(111) 2D layers grown on InSb(111)B substrates by using reflective high-energy electron diffraction (RHEED) and photoelectron spectroscopy (PES).
It is suggested that the single layer of the ST-phase Bi grows on InSb(111)B and the following Bi deposition causes layer-by-layer growth up to nominally 4 atomic layers.
Afterward, the growth behavior of Bi changes to the 3D mode with small, isolated Bi(111) islands on InSb(111)B substrates, suggested from diffused angle-resolved PES (ARPES) band dispersion.
These results support the previous observation of the fractal ST-phase Bi and provide wide-range picture of Bi 2D layer growth on lattice-matched InSb(111)B.

\section{Experimental Methods}
The InSb(111)B substrates were cleaned by repeated cycles of Ar-ion sputtering (with acceleration energy of 0.5 keV) and annealing at 540 K, resulting in In-rich (3$\times$3) surface periodicity.
Then, homo-epitaxial growth of InSb layers by co-evaporating In and Sb in Sb-rich condition with the substrate temperature of 550 K changed the substrate surface periodicity to Sb-rich (2$\times$2) as shown in the RHEED pattern (Fig. 1 (a)) \cite{Oe80}.
Bi was evaporated from an Alumina crucible to the InSb(111)B-(2$\times$2) substrates at room temperature.
During the Bi evaporation, intensities of the RHEED diffraction rods were monitored \textit{in-situ}.

Core-level PES and ARPES measurements were performed at BL5U of UVSOR-III with linearly polarized photons with the energies ranging from 40 to 120 eV.
The energy resolutions and positions of the Fermi level ($E_{\rm F}$) were calibrated by the Fermi edge of Ta foils touching the sample.
The energy resolution was evaluated to be $\sim$30 meV.
All the measurements were performed at room temperature with the photon spot size of $\sim$50 $\mu$m on the sample.

\section{Results and Discussion}
Figures 1 (a, b) show the RHEED patterns before and after the Bi growth on InSb(111)B.
As shown in Fig. 1 (b), the Bi layer after growth shows low-background and intense (1$\times$1) patterns indicating high quality of the obtained Bi film.
During the Bi growth, the surface fractional rods derived from the (2$\times$2) superlattice of the Sb-terminated InSb(111)B substrate gradually disappeared within 5 min.
We found no distinct change of the in-plane lattice constant reflected to the integer-order rods before and after the growth.
It would be because the difference of the in-plane lattice constant of 2D Bi(111) (4.53 \AA) and InSb(111) (4.58 \AA) is too small to be detected from the current RHEED diffraction rod width.

Figure 1 (c) is the intensity variations of the RHEED diffraction rods during the Bi growth.
One can find some significant peaks indicated by triangle markers, which would be derived from completion of each Bi atomic layer.
The time interval between the peaks is 3.5$\pm$0.5 min. in most cases.
The time between the start of the evaporation and the first peak is 1.5-2.0 min., different from the others.
It would be because the first peak is from the wetting layer consists not only of Bi but also of surface Sb and In atoms.
The major difference between the (0 0) and (2 0) rods is the second peak.
The second peak of (0 0) is the most intense and its peak time ($\sim$ 5 min.) is different from that of (2 0) ($\sim$6 min.).
The ST phase would not contribute to the in-plane diffraction rods other than (0 0), where all the information of the ordered surface, both in-plane and out-of-plane ones, contributes.
It is because the ST patterns do not obey the in-plane uniform-interval periodicity due to the self-similar fractal patterning.
Therefore, the intense peak appearing only at (0 0) rod would be derived from the ST phase.
After $\sim$15 min. with 5 peaks (4 layers and the wetting layer, while the 5th peak is not so significant as the former ones), the oscillation of the RHEED intensity disappears.
It suggests that the growth mode of Bi changes and the layer-by-layer growth becomes no longer valid after this period, while the in-plane periodicity of Bi(111) remains as shown in Fig. 1 (b).

Figure 2 (a) is the core-level PES spectra of Bi/InSb(111)B during the Bi growth.
The 4$d$ peaks of In and Sb decrease together with the increase of Bi 5$d$ ones.
This behavior is consistent with the growth of Bi layers capping the InSb substrate.
As indicated by the vertical dashed lines, Sb 4$d$ has tails in higher kinetic energy region at the clean substrate (0 min.), while In 4$d$ appears as a simple spin-orbital doublet.
It would be due to the surface termination layer of Sb of the Sb-rich InSb(111)B-(2$\times$2) surface.
This tail structure of Sb 4$d$ disappears with the Bi evaporation and the peak energy shifts to higher Kinetic energies.
At 6 min., both In and Sb 4$d$ levels are weak but simple doublets without tails.
This change would come from the formation of the Bi layers together with breaking the Sb-(2$\times$2) termination layers of the substrate.
As shown in Fig. 2 (b), the peak energy of the Bi 5$d$ levels also changes from the initial period (1.5 min.) to the thicker region (6.0 min. or higher).
This behavior is consistent with the RHEED oscillation suggesting the wetting-layer formation, where the surrounding elements of Bi is different from the upper layers; only the Bi atoms in the wetting layer would touch the substrate InSb.
The peak positions of the Bi 5$d$ in the 6 min. and 45 min. spectra are nearly the same, suggesting that the surrounding conditions of the Bi atoms above the wetting layer are almost identical in whole range of Bi growth.

Figure 2 (c) is the evolution of the In and Sb core-level areas normalized by corresponding Bi areas.
The core-level areas are obtained from the PES spectra by subtracting linear backgrounds estimated around each peak; for example, 55 to 61 eV for In 4$d$.
Both In and Sb peaks decay linearly in the logarithmic plot, indicating the uniform exponential decay of the core-level area.
Moreover, the slope of the linear decay changes in 12-15 min. for both In and Sb peaks, as guided by the fat lines in Fig. 2 (c).
It suggests that the growth mode of Bi changes in this region.
Combined with the disappearance of the RHEED oscillation around 15 min., it would be a change from the layer-by-layer to 3D island-type growth modes.
The latter mode has the smaller slope than that of the former mode.
This behavior can be explained by the assumed change of the growth mode, because the 3D island growth could remain some area with thin Bi thicknesses.

Figure 3 is the ARPES intensity plots of the Bi layers on InSb(111)B, measured along $\bar{\Gamma}-\bar{\rm M}$ with linearly polarized photons ($h\nu$ = 40 eV).
As guided by the white dashed lines in Figs. 3 (a, b), the hole-like bands (light and heavy holes of zinc-blende-type semiconductors) observed before Bi evaporation disappears soon after starting the Bi evaporation.
At 1.5 min. corresponding to the the wetting layer formation, no significant bands were observed.
It suggests that the wetting layer of Bi on InSb(111)B has no well-defined 2D atomic and electronic structures.
Then, at 6.0 min., where the ST-phase would appear, the new bands S1 and S2 dispersing upwards to $\bar{\Gamma}$ and downwards to $\bar{\rm M}$, respectively, appear.
However, these bands, S1 and S2, remain in the similar energy ranges in thicker regions, as shown in Figs. 3 (d, e), far above the disappearance of ST structures \cite{Liu21}.
Therefore, these bands would be from the 2D Bi(111) layers, rather than from the fractal ST surface structures.
We found no other new band structures around $E_{\rm F}$ in 4.5-6.0 min. samples.
It would be because of the same reason why the formation of the ST phase only appears in the RHEED (0 0) diffraction rod; the in-plane periods of the ST phase are not uniform and thus the atomic and electronic structures from the ST phase cannot be observed as the 2D periodic diffraction nor bands.

The binding energy of the bottom of S2 decrease together with the increasing thickness of Bi, as shown in Figs. 3 (c-f).
In the thick region (Figs. 3 (e, f)), the second feature S2' lying in deeper binding energies than S2 appears.
These can be understood by assuming S2 and S2' as the quantum-well (QW) states of Bi ultra-thin layers appearing around $\bar{\rm M}$ \cite{Hirahara06}.
Since the number of QW bands increases together with the number of the layers, one QW band, S2 in this case, should move to uniform directions during the growth so that the larger numbers of QW bands could be filled in the corresponding bulk-band projection.
This trend is consistent with the assumed layer-by-layer growth mode of the 2D Bi layers on InSb(111)B.

However, some features of the ARPES plots cannot be understood straightforwardly by the layer-by-layer growth mode.
At first, the observed band dispersion of the thick sample shown in Fig. 3 (f) is different from the bulk-like Bi(111) layers.
For example, the electron pocket forming the hexagonal Fermi contour \cite{Hirahara06, Ito16} is not observed in this work.
Secondly, the QW state appears even in the thick sample (Fig. 3(f)) where the layer-by-layer growth mode should be already broken.
These features could be explained by supposing that the surface atomic structures of the obtained 2D Bi layers have small commensurate areas.
The earlier work reported that even after the disappearing of the fractal ST phase, the surface atomic structure of Bi/InSb(111)B contains many domain walls and the commensurate 2D atomic structure is smaller than 10 atomic units \cite{Liu21}.
This would prevent the formation of the well-defined 2D surface bands as observed on the other substrates \cite{Hirahara06, Hirahara11, Yao16, Ito16}.
This hypothesis could also explain the diffused bands in whole ARPES data, comparing to the other substrates. 
With such in-plane disorder and variation of the heights of the 3D islands, the QW states from each island would be averaged as an uniform background and only the common 2D layers, which would have the maximum height among the islands in each growth period, would provide the diffused QW states to the APRES intensity plots, as Fig. 3 (f).

As shown above, the averaged 2D electronic structure of the Bi layers on InSb(111)B is diffused and it would be difficult to find out unique electronic characteristics from there.
However, at the same time, it encourages the electronic structure exploration with the better spatial resolution.
For example, the electron microscopy technique together with ARPES \cite{Matsui20} or ARPES with nano-focused photons \cite{Bostwick12} would be useful to obtain the electronic structures localized in self-similar triangle edges and corners on the ST Bi layers.
Since the majority area of the Bi layers lacks the well-defined 2D periodicity, there would be full of the 1D edges with different characteristics (lengths, corner angles, and so on).
Therefore, the Bi layers on InSb(111)B would be a fertile playground for the next-generation high-spatial resolution techniques to study the low-dimensional (0D or 1D) topological electronic structures.


\section{Summary}

In this work, we have studied the evolution of the atomic and electronic structures of the Bi(111) layers grown on InSb(111)B substrates by using RHEED and PES.
It is suggested that the single layer of the ST-phase Bi grows on InSb(111)B and the following Bi evaporation causes layer-by-layer growth up to nominally 4 atomic layers.
Afterward, the growth behavior of Bi changes to the 3D mode with small, isolated Bi(111) islands on InSb(111)B substrates suggested from diffused ARPES band dispersion.
These results support the previous observation of the fractal ST-phase Bi and provide wide-range picture of Bi 2D layer growth on lattice-matched InSb(111)B, encouraging the following researches with high spatial resolution to unveil the low-dimensional electronic structures of the Bi fractal triangles.

\section{Acknowledgement}
We acknowledge J. K. Modak, S. Ideta and K. Tanaka for their technical support during the experiments.
The core-level PES and ARPES measurements were performed under UVSOR proposals 19-563 and 20-765.
This work is supported by JSPS KAKENHI (Grants Nos. JP20K03859, JP19H01830, and JP20H04453).

\bibliography{BiRefs}

\providecommand{\noopsort}[1]{}\providecommand{\singleletter}[1]{#1}%
\begin{thebibliography}{24}%
\makeatletter
\providecommand \@ifxundefined [1]{%
 \@ifx{#1\undefined}
}%
\providecommand \@ifnum [1]{%
 \ifnum #1\expandafter \@firstoftwo
 \else \expandafter \@secondoftwo
 \fi
}%
\providecommand \@ifx [1]{%
 \ifx #1\expandafter \@firstoftwo
 \else \expandafter \@secondoftwo
 \fi
}%
\providecommand \natexlab [1]{#1}%
\providecommand \enquote  [1]{``#1''}%
\providecommand \bibnamefont  [1]{#1}%
\providecommand \bibfnamefont [1]{#1}%
\providecommand \citenamefont [1]{#1}%
\providecommand \href@noop [0]{\@secondoftwo}%
\providecommand \href [0]{\begingroup \@sanitize@url \@href}%
\providecommand \@href[1]{\@@startlink{#1}\@@href}%
\providecommand \@@href[1]{\endgroup#1\@@endlink}%
\providecommand \@sanitize@url [0]{\catcode `\\12\catcode `\$12\catcode
  `\&12\catcode `\#12\catcode `\^12\catcode `\_12\catcode `\%12\relax}%
\providecommand \@@startlink[1]{}%
\providecommand \@@endlink[0]{}%
\providecommand \url  [0]{\begingroup\@sanitize@url \@url }%
\providecommand \@url [1]{\endgroup\@href {#1}{\urlprefix }}%
\providecommand \urlprefix  [0]{URL }%
\providecommand \Eprint [0]{\href }%
\providecommand \doibase [0]{https://doi.org/}%
\providecommand \selectlanguage [0]{\@gobble}%
\providecommand \bibinfo  [0]{\@secondoftwo}%
\providecommand \bibfield  [0]{\@secondoftwo}%
\providecommand \translation [1]{[#1]}%
\providecommand \BibitemOpen [0]{}%
\providecommand \bibitemStop [0]{}%
\providecommand \bibitemNoStop [0]{.\EOS\space}%
\providecommand \EOS [0]{\spacefactor3000\relax}%
\providecommand \BibitemShut  [1]{\csname bibitem#1\endcsname}%
\let\auto@bib@innerbib\@empty
\bibitem [{\citenamefont {Hasan}\ and\ \citenamefont {Kane}(2010)}]{Hasan10}%
  \BibitemOpen
  \bibfield  {author} {\bibinfo {author} {\bibfnamefont {M.~Z.}\ \bibnamefont
  {Hasan}}\ and\ \bibinfo {author} {\bibfnamefont {C.~L.}\ \bibnamefont
  {Kane}},\ }\href {https://doi.org/10.1103/RevModPhys.82.3045} {\bibfield
  {journal} {\bibinfo  {journal} {Rev. Mod. Phys.}\ }\textbf {\bibinfo {volume}
  {82}},\ \bibinfo {pages} {3045} (\bibinfo {year} {2010})}\BibitemShut
  {NoStop}%
\bibitem [{\citenamefont {Zhang}\ \emph {et~al.}(2009)\citenamefont {Zhang},
  \citenamefont {Liu}, \citenamefont {Qi}, \citenamefont {Dai}, \citenamefont
  {Fang},\ and\ \citenamefont {Zhang}}]{Zhang09}%
  \BibitemOpen
  \bibfield  {author} {\bibinfo {author} {\bibfnamefont {H.}~\bibnamefont
  {Zhang}}, \bibinfo {author} {\bibfnamefont {C.-X.}\ \bibnamefont {Liu}},
  \bibinfo {author} {\bibfnamefont {X.-L.}\ \bibnamefont {Qi}}, \bibinfo
  {author} {\bibfnamefont {X.}~\bibnamefont {Dai}}, \bibinfo {author}
  {\bibfnamefont {Z.}~\bibnamefont {Fang}},\ and\ \bibinfo {author}
  {\bibfnamefont {S.-C.}\ \bibnamefont {Zhang}},\ }\href
  {https://doi.org/10.1038/nphys1270} {\bibfield  {journal} {\bibinfo
  {journal} {Nat. Phys.}\ }\textbf {\bibinfo {volume} {5}},\ \bibinfo {pages}
  {438} (\bibinfo {year} {2009})}\BibitemShut {NoStop}%
\bibitem [{\citenamefont {Xia}\ \emph {et~al.}(2009)\citenamefont {Xia},
  \citenamefont {Qian}, \citenamefont {Hsieh}, \citenamefont {Wray},
  \citenamefont {Pal}, \citenamefont {Lin}, \citenamefont {Bansil},
  \citenamefont {Grauer}, \citenamefont {Hor}, \citenamefont {Cava},\ and\
  \citenamefont {Hasan}}]{Xia09}%
  \BibitemOpen
  \bibfield  {author} {\bibinfo {author} {\bibfnamefont {Y.}~\bibnamefont
  {Xia}}, \bibinfo {author} {\bibfnamefont {D.}~\bibnamefont {Qian}}, \bibinfo
  {author} {\bibfnamefont {D.}~\bibnamefont {Hsieh}}, \bibinfo {author}
  {\bibfnamefont {L.}~\bibnamefont {Wray}}, \bibinfo {author} {\bibfnamefont
  {A.}~\bibnamefont {Pal}}, \bibinfo {author} {\bibfnamefont {H.}~\bibnamefont
  {Lin}}, \bibinfo {author} {\bibfnamefont {A.}~\bibnamefont {Bansil}},
  \bibinfo {author} {\bibfnamefont {D.}~\bibnamefont {Grauer}}, \bibinfo
  {author} {\bibfnamefont {Y.~S.}\ \bibnamefont {Hor}}, \bibinfo {author}
  {\bibfnamefont {R.~J.}\ \bibnamefont {Cava}},\ and\ \bibinfo {author}
  {\bibfnamefont {M.~Z.}\ \bibnamefont {Hasan}},\ }\href
  {https://doi.org/10.1038/nphys1274} {\bibfield  {journal} {\bibinfo
  {journal} {Nature physics}\ }\textbf {\bibinfo {volume} {5}},\ \bibinfo
  {pages} {398} (\bibinfo {year} {2009})}\BibitemShut {NoStop}%
\bibitem [{\citenamefont {Hsieh}\ \emph {et~al.}(2008)\citenamefont {Hsieh},
  \citenamefont {Qian}, \citenamefont {Wray}, \citenamefont {Xia},
  \citenamefont {Hor}, \citenamefont {Cava},\ and\ \citenamefont
  {Hasan}}]{Hsieh08}%
  \BibitemOpen
  \bibfield  {author} {\bibinfo {author} {\bibfnamefont {D.}~\bibnamefont
  {Hsieh}}, \bibinfo {author} {\bibfnamefont {D.}~\bibnamefont {Qian}},
  \bibinfo {author} {\bibfnamefont {L.}~\bibnamefont {Wray}}, \bibinfo {author}
  {\bibfnamefont {Y.}~\bibnamefont {Xia}}, \bibinfo {author} {\bibfnamefont
  {Y.~S.}\ \bibnamefont {Hor}}, \bibinfo {author} {\bibfnamefont {R.~J.}\
  \bibnamefont {Cava}},\ and\ \bibinfo {author} {\bibfnamefont {M.~Z.}\
  \bibnamefont {Hasan}},\ }\href {https://doi.org/10.1038/nature06843}
  {\bibfield  {journal} {\bibinfo  {journal} {Nature}\ }\textbf {\bibinfo
  {volume} {452}},\ \bibinfo {pages} {970} (\bibinfo {year}
  {2008})}\BibitemShut {NoStop}%
\bibitem [{\citenamefont {Murakami}(2006)}]{Murakami06}%
  \BibitemOpen
  \bibfield  {author} {\bibinfo {author} {\bibfnamefont {S.}~\bibnamefont
  {Murakami}},\ }\href {https://doi.org/10.1103/PhysRevLett.97.236805}
  {\bibfield  {journal} {\bibinfo  {journal} {Phys. Rev. Lett.}\ }\textbf
  {\bibinfo {volume} {97}},\ \bibinfo {pages} {236805} (\bibinfo {year}
  {2006})}\BibitemShut {NoStop}%
\bibitem [{\citenamefont {Liu}\ \emph {et~al.}(2011)\citenamefont {Liu},
  \citenamefont {Liu}, \citenamefont {Wu}, \citenamefont {Duan}, \citenamefont
  {Liu},\ and\ \citenamefont {Wu}}]{Liu11}%
  \BibitemOpen
  \bibfield  {author} {\bibinfo {author} {\bibfnamefont {Z.}~\bibnamefont
  {Liu}}, \bibinfo {author} {\bibfnamefont {C.-X.}\ \bibnamefont {Liu}},
  \bibinfo {author} {\bibfnamefont {Y.-S.}\ \bibnamefont {Wu}}, \bibinfo
  {author} {\bibfnamefont {W.-H.}\ \bibnamefont {Duan}}, \bibinfo {author}
  {\bibfnamefont {F.}~\bibnamefont {Liu}},\ and\ \bibinfo {author}
  {\bibfnamefont {J.}~\bibnamefont {Wu}},\ }\href
  {https://doi.org/10.1103/PhysRevLett.107.136805} {\bibfield  {journal}
  {\bibinfo  {journal} {Phys. Rev. Lett.}\ }\textbf {\bibinfo {volume} {107}},\
  \bibinfo {pages} {136805} (\bibinfo {year} {2011})}\BibitemShut {NoStop}%
\bibitem [{\citenamefont {Huang}\ \emph {et~al.}(2013)\citenamefont {Huang},
  \citenamefont {Chuang}, \citenamefont {Hsu}, \citenamefont {Liu},
  \citenamefont {Chang}, \citenamefont {Lin},\ and\ \citenamefont
  {Bansil}}]{Huang13}%
  \BibitemOpen
  \bibfield  {author} {\bibinfo {author} {\bibfnamefont {Z.-Q.}\ \bibnamefont
  {Huang}}, \bibinfo {author} {\bibfnamefont {F.-C.}\ \bibnamefont {Chuang}},
  \bibinfo {author} {\bibfnamefont {C.-H.}\ \bibnamefont {Hsu}}, \bibinfo
  {author} {\bibfnamefont {Y.-T.}\ \bibnamefont {Liu}}, \bibinfo {author}
  {\bibfnamefont {H.-R.}\ \bibnamefont {Chang}}, \bibinfo {author}
  {\bibfnamefont {H.}~\bibnamefont {Lin}},\ and\ \bibinfo {author}
  {\bibfnamefont {A.}~\bibnamefont {Bansil}},\ }\href
  {https://doi.org/10.1103/PhysRevB.88.165301} {\bibfield  {journal} {\bibinfo
  {journal} {Phys. Rev. B}\ }\textbf {\bibinfo {volume} {88}},\ \bibinfo
  {pages} {165301} (\bibinfo {year} {2013})}\BibitemShut {NoStop}%
\bibitem [{\citenamefont {Kadioglu}\ \emph {et~al.}(2017)\citenamefont
  {Kadioglu}, \citenamefont {Kilic}, \citenamefont {Demirci}, \citenamefont
  {Akt\"urk}, \citenamefont {Akt\"urk},\ and\ \citenamefont
  {Ciraci}}]{Kadioglu17}%
  \BibitemOpen
  \bibfield  {author} {\bibinfo {author} {\bibfnamefont {Y.}~\bibnamefont
  {Kadioglu}}, \bibinfo {author} {\bibfnamefont {S.~B.}\ \bibnamefont {Kilic}},
  \bibinfo {author} {\bibfnamefont {S.}~\bibnamefont {Demirci}}, \bibinfo
  {author} {\bibfnamefont {O.~U.}\ \bibnamefont {Akt\"urk}}, \bibinfo {author}
  {\bibfnamefont {E.}~\bibnamefont {Akt\"urk}},\ and\ \bibinfo {author}
  {\bibfnamefont {S.}~\bibnamefont {Ciraci}},\ }\href
  {https://doi.org/10.1103/PhysRevB.96.245424} {\bibfield  {journal} {\bibinfo
  {journal} {Phys. Rev. B}\ }\textbf {\bibinfo {volume} {96}},\ \bibinfo
  {pages} {245424} (\bibinfo {year} {2017})}\BibitemShut {NoStop}%
\bibitem [{\citenamefont {Liu}\ \emph {et~al.}(2017)\citenamefont {Liu},
  \citenamefont {Huang}, \citenamefont {Chen}, \citenamefont {Li},
  \citenamefont {Cao},\ and\ \citenamefont {He}}]{Liu17}%
  \BibitemOpen
  \bibfield  {author} {\bibinfo {author} {\bibfnamefont {M.-Y.}\ \bibnamefont
  {Liu}}, \bibinfo {author} {\bibfnamefont {Y.}~\bibnamefont {Huang}}, \bibinfo
  {author} {\bibfnamefont {Q.-Y.}\ \bibnamefont {Chen}}, \bibinfo {author}
  {\bibfnamefont {Z.-Y.}\ \bibnamefont {Li}}, \bibinfo {author} {\bibfnamefont
  {C.}~\bibnamefont {Cao}},\ and\ \bibinfo {author} {\bibfnamefont
  {Y.}~\bibnamefont {He}},\ }\href {https://doi.org/10.1039/C7RA05787C}
  {\bibfield  {journal} {\bibinfo  {journal} {RSC Adv.}\ }\textbf {\bibinfo
  {volume} {7}},\ \bibinfo {pages} {39546} (\bibinfo {year}
  {2017})}\BibitemShut {NoStop}%
\bibitem [{\citenamefont {Nagao}\ \emph {et~al.}(2004)\citenamefont {Nagao},
  \citenamefont {Sadowski}, \citenamefont {Saito}, \citenamefont {Yaginuma},
  \citenamefont {Fujikawa}, \citenamefont {Kogure}, \citenamefont {Ohno},
  \citenamefont {Hasegawa}, \citenamefont {Hasegawa},\ and\ \citenamefont
  {Sakurai}}]{Nagao04}%
  \BibitemOpen
  \bibfield  {author} {\bibinfo {author} {\bibfnamefont {T.}~\bibnamefont
  {Nagao}}, \bibinfo {author} {\bibfnamefont {J.~T.}\ \bibnamefont {Sadowski}},
  \bibinfo {author} {\bibfnamefont {M.}~\bibnamefont {Saito}}, \bibinfo
  {author} {\bibfnamefont {S.}~\bibnamefont {Yaginuma}}, \bibinfo {author}
  {\bibfnamefont {Y.}~\bibnamefont {Fujikawa}}, \bibinfo {author}
  {\bibfnamefont {T.}~\bibnamefont {Kogure}}, \bibinfo {author} {\bibfnamefont
  {T.}~\bibnamefont {Ohno}}, \bibinfo {author} {\bibfnamefont {Y.}~\bibnamefont
  {Hasegawa}}, \bibinfo {author} {\bibfnamefont {S.}~\bibnamefont {Hasegawa}},\
  and\ \bibinfo {author} {\bibfnamefont {T.}~\bibnamefont {Sakurai}},\ }\href
  {https://doi.org/10.1103/PhysRevLett.93.105501} {\bibfield  {journal}
  {\bibinfo  {journal} {Phys. Rev. Lett.}\ }\textbf {\bibinfo {volume} {93}},\
  \bibinfo {pages} {105501} (\bibinfo {year} {2004})}\BibitemShut {NoStop}%
\bibitem [{\citenamefont {Hirahara}\ \emph {et~al.}(2006)\citenamefont
  {Hirahara}, \citenamefont {Nagao}, \citenamefont {Matsuda}, \citenamefont
  {Bihlmayer}, \citenamefont {Chulkov}, \citenamefont {Koroteev}, \citenamefont
  {Echenique}, \citenamefont {Saito},\ and\ \citenamefont
  {Hasegawa}}]{Hirahara06}%
  \BibitemOpen
  \bibfield  {author} {\bibinfo {author} {\bibfnamefont {T.}~\bibnamefont
  {Hirahara}}, \bibinfo {author} {\bibfnamefont {T.}~\bibnamefont {Nagao}},
  \bibinfo {author} {\bibfnamefont {I.}~\bibnamefont {Matsuda}}, \bibinfo
  {author} {\bibfnamefont {G.}~\bibnamefont {Bihlmayer}}, \bibinfo {author}
  {\bibfnamefont {E.~V.}\ \bibnamefont {Chulkov}}, \bibinfo {author}
  {\bibfnamefont {Y.~M.}\ \bibnamefont {Koroteev}}, \bibinfo {author}
  {\bibfnamefont {P.~M.}\ \bibnamefont {Echenique}}, \bibinfo {author}
  {\bibfnamefont {M.}~\bibnamefont {Saito}},\ and\ \bibinfo {author}
  {\bibfnamefont {S.}~\bibnamefont {Hasegawa}},\ }\href
  {https://doi.org/10.1103/PhysRevLett.97.146803} {\bibfield  {journal}
  {\bibinfo  {journal} {Phys. Rev. Lett.}\ }\textbf {\bibinfo {volume} {97}},\
  \bibinfo {pages} {146803} (\bibinfo {year} {2006})}\BibitemShut {NoStop}%
\bibitem [{\citenamefont {Hatta}\ \emph {et~al.}(2009)\citenamefont {Hatta},
  \citenamefont {Ohtsubo}, \citenamefont {Miyamoto}, \citenamefont {Okuyama},\
  and\ \citenamefont {Aruga}}]{Hatta09}%
  \BibitemOpen
  \bibfield  {author} {\bibinfo {author} {\bibfnamefont {S.}~\bibnamefont
  {Hatta}}, \bibinfo {author} {\bibfnamefont {Y.}~\bibnamefont {Ohtsubo}},
  \bibinfo {author} {\bibfnamefont {S.}~\bibnamefont {Miyamoto}}, \bibinfo
  {author} {\bibfnamefont {H.}~\bibnamefont {Okuyama}},\ and\ \bibinfo {author}
  {\bibfnamefont {T.}~\bibnamefont {Aruga}},\ }\href
  {https://doi.org/https://doi.org/10.1016/j.apsusc.2009.05.079} {\bibfield
  {journal} {\bibinfo  {journal} {Applied Surface Science}\ }\textbf {\bibinfo
  {volume} {256}},\ \bibinfo {pages} {1252} (\bibinfo {year}
  {2009})}\BibitemShut {NoStop}%
\bibitem [{\citenamefont {Ito}\ \emph {et~al.}(2016)\citenamefont {Ito},
  \citenamefont {Feng}, \citenamefont {Arita}, \citenamefont {Takayama},
  \citenamefont {Liu}, \citenamefont {Someya}, \citenamefont {Chen},
  \citenamefont {Iimori}, \citenamefont {Namatame}, \citenamefont {Taniguchi},
  \citenamefont {Cheng}, \citenamefont {Tang}, \citenamefont {Komori},
  \citenamefont {Kobayashi}, \citenamefont {Chiang},\ and\ \citenamefont
  {Matsuda}}]{Ito16}%
  \BibitemOpen
  \bibfield  {author} {\bibinfo {author} {\bibfnamefont {S.}~\bibnamefont
  {Ito}}, \bibinfo {author} {\bibfnamefont {B.}~\bibnamefont {Feng}}, \bibinfo
  {author} {\bibfnamefont {M.}~\bibnamefont {Arita}}, \bibinfo {author}
  {\bibfnamefont {A.}~\bibnamefont {Takayama}}, \bibinfo {author}
  {\bibfnamefont {R.-Y.}\ \bibnamefont {Liu}}, \bibinfo {author} {\bibfnamefont
  {T.}~\bibnamefont {Someya}}, \bibinfo {author} {\bibfnamefont {W.-C.}\
  \bibnamefont {Chen}}, \bibinfo {author} {\bibfnamefont {T.}~\bibnamefont
  {Iimori}}, \bibinfo {author} {\bibfnamefont {H.}~\bibnamefont {Namatame}},
  \bibinfo {author} {\bibfnamefont {M.}~\bibnamefont {Taniguchi}}, \bibinfo
  {author} {\bibfnamefont {C.-M.}\ \bibnamefont {Cheng}}, \bibinfo {author}
  {\bibfnamefont {S.-J.}\ \bibnamefont {Tang}}, \bibinfo {author}
  {\bibfnamefont {F.}~\bibnamefont {Komori}}, \bibinfo {author} {\bibfnamefont
  {K.}~\bibnamefont {Kobayashi}}, \bibinfo {author} {\bibfnamefont {T.-C.}\
  \bibnamefont {Chiang}},\ and\ \bibinfo {author} {\bibfnamefont
  {I.}~\bibnamefont {Matsuda}},\ }\href
  {https://doi.org/10.1103/PhysRevLett.117.236402} {\bibfield  {journal}
  {\bibinfo  {journal} {Phys. Rev. Lett.}\ }\textbf {\bibinfo {volume} {117}},\
  \bibinfo {pages} {236402} (\bibinfo {year} {2016})}\BibitemShut {NoStop}%
\bibitem [{\citenamefont {Hirahara}\ \emph {et~al.}(2011)\citenamefont
  {Hirahara}, \citenamefont {Bihlmayer}, \citenamefont {Sakamoto},
  \citenamefont {Yamada}, \citenamefont {Miyazaki}, \citenamefont {Kimura},
  \citenamefont {Bl\"ugel},\ and\ \citenamefont {Hasegawa}}]{Hirahara11}%
  \BibitemOpen
  \bibfield  {author} {\bibinfo {author} {\bibfnamefont {T.}~\bibnamefont
  {Hirahara}}, \bibinfo {author} {\bibfnamefont {G.}~\bibnamefont {Bihlmayer}},
  \bibinfo {author} {\bibfnamefont {Y.}~\bibnamefont {Sakamoto}}, \bibinfo
  {author} {\bibfnamefont {M.}~\bibnamefont {Yamada}}, \bibinfo {author}
  {\bibfnamefont {H.}~\bibnamefont {Miyazaki}}, \bibinfo {author}
  {\bibfnamefont {S.-i.}\ \bibnamefont {Kimura}}, \bibinfo {author}
  {\bibfnamefont {S.}~\bibnamefont {Bl\"ugel}},\ and\ \bibinfo {author}
  {\bibfnamefont {S.}~\bibnamefont {Hasegawa}},\ }\href
  {https://doi.org/10.1103/PhysRevLett.107.166801} {\bibfield  {journal}
  {\bibinfo  {journal} {Phys. Rev. Lett.}\ }\textbf {\bibinfo {volume} {107}},\
  \bibinfo {pages} {166801} (\bibinfo {year} {2011})}\BibitemShut {NoStop}%
\bibitem [{\citenamefont {Hirahara}\ \emph {et~al.}(2012)\citenamefont
  {Hirahara}, \citenamefont {Fukui}, \citenamefont {Shirasawa}, \citenamefont
  {Yamada}, \citenamefont {Aitani}, \citenamefont {Miyazaki}, \citenamefont
  {Matsunami}, \citenamefont {Kimura}, \citenamefont {Takahashi}, \citenamefont
  {Hasegawa},\ and\ \citenamefont {Kobayashi}}]{Hirahara12}%
  \BibitemOpen
  \bibfield  {author} {\bibinfo {author} {\bibfnamefont {T.}~\bibnamefont
  {Hirahara}}, \bibinfo {author} {\bibfnamefont {N.}~\bibnamefont {Fukui}},
  \bibinfo {author} {\bibfnamefont {T.}~\bibnamefont {Shirasawa}}, \bibinfo
  {author} {\bibfnamefont {M.}~\bibnamefont {Yamada}}, \bibinfo {author}
  {\bibfnamefont {M.}~\bibnamefont {Aitani}}, \bibinfo {author} {\bibfnamefont
  {H.}~\bibnamefont {Miyazaki}}, \bibinfo {author} {\bibfnamefont
  {M.}~\bibnamefont {Matsunami}}, \bibinfo {author} {\bibfnamefont
  {S.}~\bibnamefont {Kimura}}, \bibinfo {author} {\bibfnamefont
  {T.}~\bibnamefont {Takahashi}}, \bibinfo {author} {\bibfnamefont
  {S.}~\bibnamefont {Hasegawa}},\ and\ \bibinfo {author} {\bibfnamefont
  {K.}~\bibnamefont {Kobayashi}},\ }\href
  {https://doi.org/10.1103/PhysRevLett.109.227401} {\bibfield  {journal}
  {\bibinfo  {journal} {Phys. Rev. Lett.}\ }\textbf {\bibinfo {volume} {109}},\
  \bibinfo {pages} {227401} (\bibinfo {year} {2012})}\BibitemShut {NoStop}%
\bibitem [{\citenamefont {Chen}\ \emph {et~al.}(2012)\citenamefont {Chen},
  \citenamefont {Peng}, \citenamefont {Zhang}, \citenamefont {Wang},
  \citenamefont {He}, \citenamefont {Ma},\ and\ \citenamefont {Xue}}]{Chen12}%
  \BibitemOpen
  \bibfield  {author} {\bibinfo {author} {\bibfnamefont {M.}~\bibnamefont
  {Chen}}, \bibinfo {author} {\bibfnamefont {J.-P.}\ \bibnamefont {Peng}},
  \bibinfo {author} {\bibfnamefont {H.-M.}\ \bibnamefont {Zhang}}, \bibinfo
  {author} {\bibfnamefont {L.-L.}\ \bibnamefont {Wang}}, \bibinfo {author}
  {\bibfnamefont {K.}~\bibnamefont {He}}, \bibinfo {author} {\bibfnamefont
  {X.-C.}\ \bibnamefont {Ma}},\ and\ \bibinfo {author} {\bibfnamefont {Q.-K.}\
  \bibnamefont {Xue}},\ }\href {https://doi.org/10.1063/1.4747715} {\bibfield
  {journal} {\bibinfo  {journal} {Applied Physics Letters}\ }\textbf {\bibinfo
  {volume} {101}},\ \bibinfo {pages} {081603} (\bibinfo {year}
  {2012})}\BibitemShut {NoStop}%
\bibitem [{\citenamefont {Yao}\ \emph {et~al.}(2016)\citenamefont {Yao},
  \citenamefont {Zhu}, \citenamefont {Han}, \citenamefont {Guan}, \citenamefont
  {Liu}, \citenamefont {Qian},\ and\ \citenamefont {Jia}}]{Yao16}%
  \BibitemOpen
  \bibfield  {author} {\bibinfo {author} {\bibfnamefont {M.-Y.}\ \bibnamefont
  {Yao}}, \bibinfo {author} {\bibfnamefont {F.}~\bibnamefont {Zhu}}, \bibinfo
  {author} {\bibfnamefont {C.}~\bibnamefont {Han}}, \bibinfo {author}
  {\bibfnamefont {D.}~\bibnamefont {Guan}}, \bibinfo {author} {\bibfnamefont
  {C.}~\bibnamefont {Liu}}, \bibinfo {author} {\bibfnamefont {D.}~\bibnamefont
  {Qian}},\ and\ \bibinfo {author} {\bibfnamefont {J.-f.}\ \bibnamefont
  {Jia}},\ }\href@noop {} {\bibfield  {journal} {\bibinfo  {journal}
  {Scientific reports}\ }\textbf {\bibinfo {volume} {6}},\ \bibinfo {pages} {1}
  (\bibinfo {year} {2016})}\BibitemShut {NoStop}%
\bibitem [{\citenamefont {Reis}\ \emph {et~al.}(2017)\citenamefont {Reis},
  \citenamefont {Li}, \citenamefont {Dudy}, \citenamefont {Bauernfeind},
  \citenamefont {Glass}, \citenamefont {Hanke}, \citenamefont {Thomale},
  \citenamefont {Schäfer},\ and\ \citenamefont {Claessen}}]{Reis17}%
  \BibitemOpen
  \bibfield  {author} {\bibinfo {author} {\bibfnamefont {F.}~\bibnamefont
  {Reis}}, \bibinfo {author} {\bibfnamefont {G.}~\bibnamefont {Li}}, \bibinfo
  {author} {\bibfnamefont {L.}~\bibnamefont {Dudy}}, \bibinfo {author}
  {\bibfnamefont {M.}~\bibnamefont {Bauernfeind}}, \bibinfo {author}
  {\bibfnamefont {S.}~\bibnamefont {Glass}}, \bibinfo {author} {\bibfnamefont
  {W.}~\bibnamefont {Hanke}}, \bibinfo {author} {\bibfnamefont
  {R.}~\bibnamefont {Thomale}}, \bibinfo {author} {\bibfnamefont
  {J.}~\bibnamefont {Schäfer}},\ and\ \bibinfo {author} {\bibfnamefont
  {R.}~\bibnamefont {Claessen}},\ }\href
  {https://doi.org/10.1126/science.aai8142} {\bibfield  {journal} {\bibinfo
  {journal} {Science}\ }\textbf {\bibinfo {volume} {357}},\ \bibinfo {pages}
  {287} (\bibinfo {year} {2017})}\BibitemShut {NoStop}%
\bibitem [{\citenamefont {Nicolaï}\ \emph {et~al.}(2019)\citenamefont
  {Nicolaï}, \citenamefont {Mariot}, \citenamefont {Djukic}, \citenamefont
  {Wang}, \citenamefont {Heckmann}, \citenamefont {Richter}, \citenamefont
  {Kanski}, \citenamefont {Leandersson}, \citenamefont {Balasubramanian},
  \citenamefont {Sadowski}, \citenamefont {Braun}, \citenamefont {Ebert},
  \citenamefont {Vobornik}, \citenamefont {Fujii}, \citenamefont
  {Min{\'{a}}r},\ and\ \citenamefont {Hricovini}}]{Nicolai19}%
  \BibitemOpen
  \bibfield  {author} {\bibinfo {author} {\bibfnamefont {L.}~\bibnamefont
  {Nicolaï}}, \bibinfo {author} {\bibfnamefont {J.-M.}\ \bibnamefont
  {Mariot}}, \bibinfo {author} {\bibfnamefont {U.}~\bibnamefont {Djukic}},
  \bibinfo {author} {\bibfnamefont {W.}~\bibnamefont {Wang}}, \bibinfo {author}
  {\bibfnamefont {O.}~\bibnamefont {Heckmann}}, \bibinfo {author}
  {\bibfnamefont {M.~C.}\ \bibnamefont {Richter}}, \bibinfo {author}
  {\bibfnamefont {J.}~\bibnamefont {Kanski}}, \bibinfo {author} {\bibfnamefont
  {M.}~\bibnamefont {Leandersson}}, \bibinfo {author} {\bibfnamefont
  {T.}~\bibnamefont {Balasubramanian}}, \bibinfo {author} {\bibfnamefont
  {J.}~\bibnamefont {Sadowski}}, \bibinfo {author} {\bibfnamefont
  {J.}~\bibnamefont {Braun}}, \bibinfo {author} {\bibfnamefont
  {H.}~\bibnamefont {Ebert}}, \bibinfo {author} {\bibfnamefont
  {I.}~\bibnamefont {Vobornik}}, \bibinfo {author} {\bibfnamefont
  {J.}~\bibnamefont {Fujii}}, \bibinfo {author} {\bibfnamefont
  {J.}~\bibnamefont {Min{\'{a}}r}},\ and\ \bibinfo {author} {\bibfnamefont
  {K.}~\bibnamefont {Hricovini}},\ }\href
  {https://doi.org/10.1088/1367-2630/ab5c14} {\bibfield  {journal} {\bibinfo
  {journal} {New J. Phys.}\ }\textbf {\bibinfo {volume} {21}},\ \bibinfo
  {pages} {123012} (\bibinfo {year} {2019})}\BibitemShut {NoStop}%
\bibitem [{\citenamefont {Liu}\ \emph {et~al.}(2021)\citenamefont {Liu},
  \citenamefont {Zhou}, \citenamefont {Wang}, \citenamefont {Yin},
  \citenamefont {Li}, \citenamefont {Huang}, \citenamefont {Guan},
  \citenamefont {Li}, \citenamefont {Wang}, \citenamefont {Zheng},
  \citenamefont {Liu}, \citenamefont {Han}, \citenamefont {Evans},
  \citenamefont {Liu},\ and\ \citenamefont {Jia}}]{Liu21}%
  \BibitemOpen
  \bibfield  {author} {\bibinfo {author} {\bibfnamefont {C.}~\bibnamefont
  {Liu}}, \bibinfo {author} {\bibfnamefont {Y.}~\bibnamefont {Zhou}}, \bibinfo
  {author} {\bibfnamefont {G.}~\bibnamefont {Wang}}, \bibinfo {author}
  {\bibfnamefont {Y.}~\bibnamefont {Yin}}, \bibinfo {author} {\bibfnamefont
  {C.}~\bibnamefont {Li}}, \bibinfo {author} {\bibfnamefont {H.}~\bibnamefont
  {Huang}}, \bibinfo {author} {\bibfnamefont {D.}~\bibnamefont {Guan}},
  \bibinfo {author} {\bibfnamefont {Y.}~\bibnamefont {Li}}, \bibinfo {author}
  {\bibfnamefont {S.}~\bibnamefont {Wang}}, \bibinfo {author} {\bibfnamefont
  {H.}~\bibnamefont {Zheng}}, \bibinfo {author} {\bibfnamefont
  {C.}~\bibnamefont {Liu}}, \bibinfo {author} {\bibfnamefont {Y.}~\bibnamefont
  {Han}}, \bibinfo {author} {\bibfnamefont {J.~W.}\ \bibnamefont {Evans}},
  \bibinfo {author} {\bibfnamefont {F.}~\bibnamefont {Liu}},\ and\ \bibinfo
  {author} {\bibfnamefont {J.}~\bibnamefont {Jia}},\ }\href
  {https://doi.org/10.1103/PhysRevLett.126.176102} {\bibfield  {journal}
  {\bibinfo  {journal} {Phys. Rev. Lett.}\ }\textbf {\bibinfo {volume} {126}},\
  \bibinfo {pages} {176102} (\bibinfo {year} {2021})}\BibitemShut {NoStop}%
\bibitem [{\citenamefont {Sierpi\'{n}ski}(1915)}]{Sierpinski15}%
  \BibitemOpen
  \bibfield  {author} {\bibinfo {author} {\bibfnamefont {W.}~\bibnamefont
  {Sierpi\'{n}ski}},\ }\href@noop {} {\bibfield  {journal} {\bibinfo  {journal}
  {CR Acad. Sci}\ }\textbf {\bibinfo {volume} {160}},\ \bibinfo {pages} {162}
  (\bibinfo {year} {1915})}\BibitemShut {NoStop}%
\bibitem [{\citenamefont {Oe}\ \emph {et~al.}(1980)\citenamefont {Oe},
  \citenamefont {Ando},\ and\ \citenamefont {Sugiyama}}]{Oe80}%
  \BibitemOpen
  \bibfield  {author} {\bibinfo {author} {\bibfnamefont {K.}~\bibnamefont
  {Oe}}, \bibinfo {author} {\bibfnamefont {S.}~\bibnamefont {Ando}},\ and\
  \bibinfo {author} {\bibfnamefont {K.}~\bibnamefont {Sugiyama}},\ }\href
  {https://doi.org/10.1143/jjap.19.l417} {\bibfield  {journal} {\bibinfo
  {journal} {Jpn. J. Appl. Phys.}\ }\textbf {\bibinfo {volume} {19}},\ \bibinfo
  {pages} {L417} (\bibinfo {year} {1980})}\BibitemShut {NoStop}%
\bibitem [{\citenamefont {Matsui}\ \emph {et~al.}(2020)\citenamefont {Matsui},
  \citenamefont {Makita}, \citenamefont {Matsuda}, \citenamefont {Yano},
  \citenamefont {Nakamura}, \citenamefont {Tanaka}, \citenamefont {Suga},\ and\
  \citenamefont {Kera}}]{Matsui20}%
  \BibitemOpen
  \bibfield  {author} {\bibinfo {author} {\bibfnamefont {F.}~\bibnamefont
  {Matsui}}, \bibinfo {author} {\bibfnamefont {S.}~\bibnamefont {Makita}},
  \bibinfo {author} {\bibfnamefont {H.}~\bibnamefont {Matsuda}}, \bibinfo
  {author} {\bibfnamefont {T.}~\bibnamefont {Yano}}, \bibinfo {author}
  {\bibfnamefont {E.}~\bibnamefont {Nakamura}}, \bibinfo {author}
  {\bibfnamefont {K.}~\bibnamefont {Tanaka}}, \bibinfo {author} {\bibfnamefont
  {S.}~\bibnamefont {Suga}},\ and\ \bibinfo {author} {\bibfnamefont
  {S.}~\bibnamefont {Kera}},\ }\href
  {https://doi.org/10.35848/1347-4065/ab9184} {\bibfield  {journal} {\bibinfo
  {journal} {Jpn. J. Appl. Phys.}\ }\textbf {\bibinfo {volume} {59}},\ \bibinfo
  {pages} {067001} (\bibinfo {year} {2020})}\BibitemShut {NoStop}%
\bibitem [{\citenamefont {Bostwick}\ \emph {et~al.}(2012)\citenamefont
  {Bostwick}, \citenamefont {Rotenberg}, \citenamefont {Avila},\ and\
  \citenamefont {Asensio}}]{Bostwick12}%
  \BibitemOpen
  \bibfield  {author} {\bibinfo {author} {\bibfnamefont {A.}~\bibnamefont
  {Bostwick}}, \bibinfo {author} {\bibfnamefont {E.}~\bibnamefont {Rotenberg}},
  \bibinfo {author} {\bibfnamefont {J.}~\bibnamefont {Avila}},\ and\ \bibinfo
  {author} {\bibfnamefont {M.~C.}\ \bibnamefont {Asensio}},\ }\href@noop {}
  {\bibfield  {journal} {\bibinfo  {journal} {Synchrotron radiation news}\
  }\textbf {\bibinfo {volume} {25}},\ \bibinfo {pages} {19} (\bibinfo {year}
  {2012})}\BibitemShut {NoStop}%
\end{thebibliography}%




\begin{figure}[p]
\includegraphics[width=80mm]{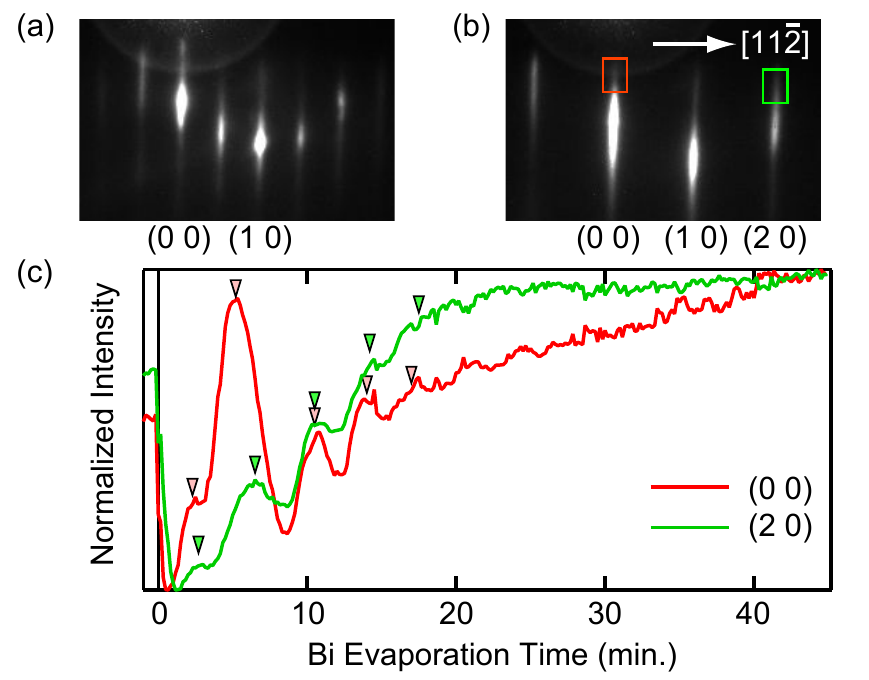}
\caption{\label{figure 1}
(a, b) RHEED patterns of (a) the InSb(111)B-(2$\times$2) substrate and (b) Bi grown on InSb(111)B with the evaporation time of 45 min. Electron energy is 12.5 keV. Rectangles in (b) represent the area where the RHEED intensities were monitored.
(c) RHEED intensity oscillations during Bi growth on InSb(111)B-(2$\times$2). Triangle markers indicate the peak positions of each diffraction rod.
}
\end{figure}

\begin{figure}[p]
\includegraphics[width=80mm]{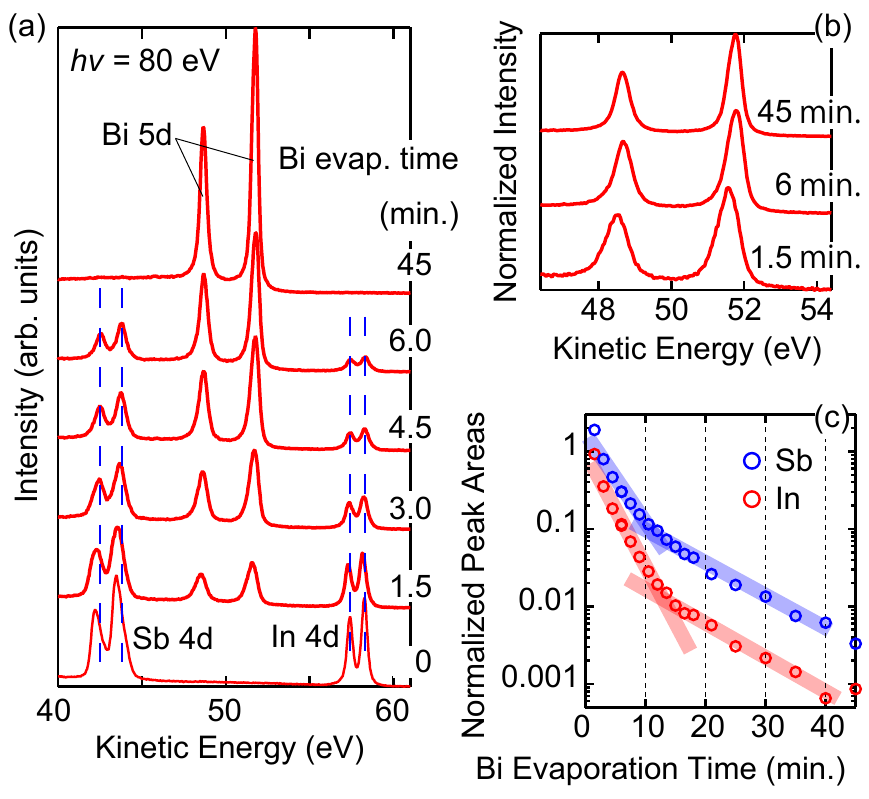}
\caption{\label{figure 2}
(a) Core-level PES spectra upon Bi evaporation on InSb(111)B.
Vertical dashed lines are guides to the eye for peak positions of In 4$d$ and Sb 4$d$ levels.
(b) Close-up spectra of Bi 5$d$ levels from the same data as (a). Each spectra is normalized by the maximum values in the kinetic energy range shown here (46 to 54.5 eV).
(c) Evolution of In and Sb core-level areas normalized by corresponding Bi core-level areas.
}
\end{figure}

\begin{figure}[p]
\includegraphics[width=80mm]{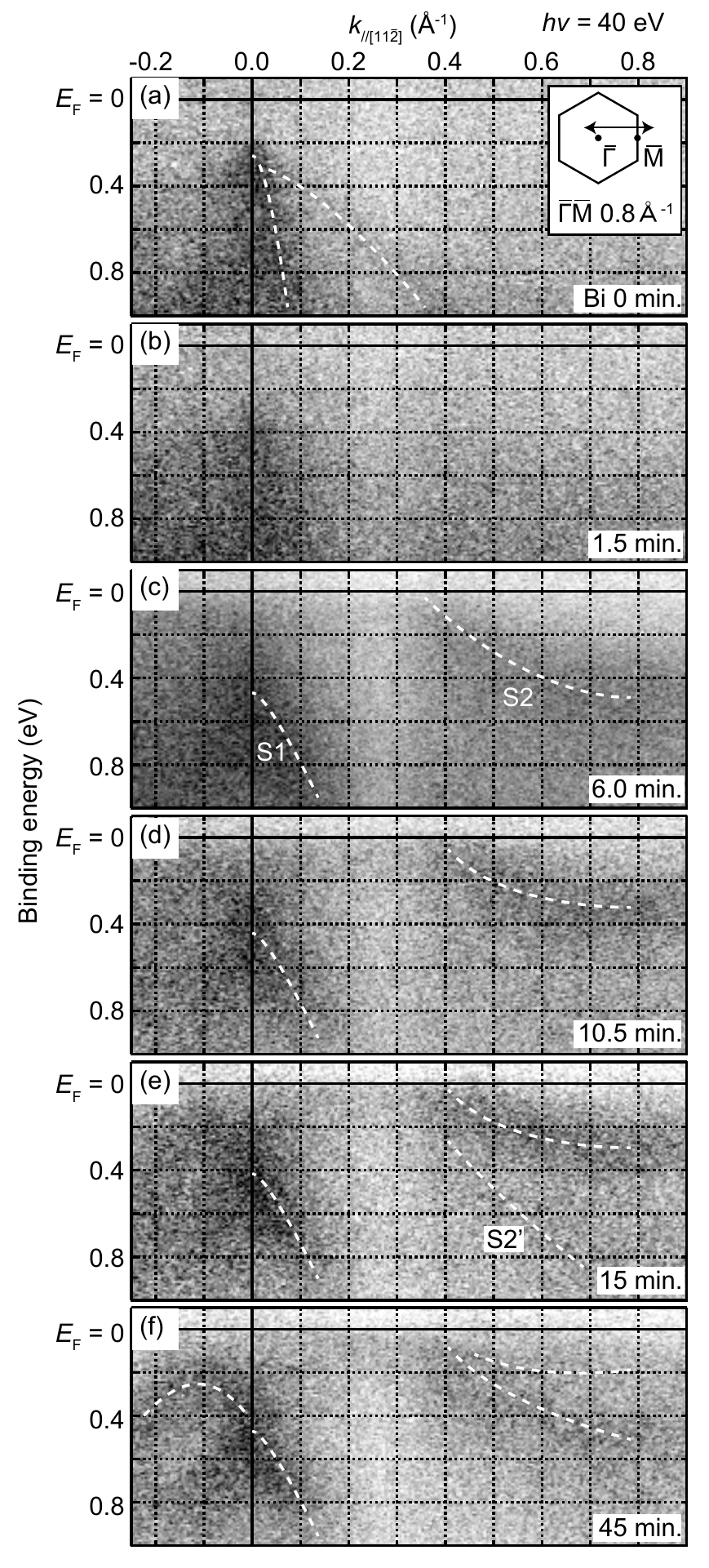}
\caption{\label{figure 3}
ARPES intensity plots obtained from Bi layers grown on InSb(111)B along $\bar{\Gamma}-\bar{\rm M}$ (parallel to [11$\bar{2}$]).
The inset in (a) depicts the $k_{//[11\bar{2}]}$ region observed.
Dashed lines guide the observed band dispersion.
}
\end{figure}

\end{document}